\documentclass[aps,showpacs,twocolumn]{revtex4}

\usepackage{graphicx}
\usepackage{bm}
\usepackage{amsmath}
\usepackage{color}

\newcommand{\be}{\begin{eqnarray}}
\newcommand{\ee}{\end{eqnarray}}

 \newcommand{\gsim}{\mathrel{\hbox{\rlap{\lower.55ex \hbox {$\sim$}}
                   \kern-.3em \raise.4ex \hbox{$>$}}}}
\newcommand{\lsim}{\mathrel{\hbox{\rlap{\lower.55ex \hbox {$\sim$}}
                   \kern-.3em \raise.4ex \hbox{$<$}}}}

\newcommand{\ba}{\begin{eqnarray}}
\newcommand{\ea}{\end{eqnarray}}

\newcommand{\snn}{\sqrt {s_{NN}}}

\begin{document}


\title{Rotating quark-gluon plasma in relativistic heavy ion collisions}

\author{Yin Jiang$^{1}$, Zi-Wei Lin$^2$, and Jinfeng Liao$^{1,3}$}
\address{$^1$ Physics Department and Center for Exploration of Energy and Matter,
Indiana University, 2401 N Milo B. Sampson Lane, Bloomington, IN 47408, USA.\\
$^2$ Department of Physics, East Carolina University, Greenville, NC 27858, USA. \\
$^3$ RIKEN BNL Research Center, Bldg. 510A, Brookhaven National Laboratory, Upton, NY 11973, USA.}
\date{\today}

\begin{abstract}
We study the rotational collective motion of the quark-gluon plasma in relativistic heavy ion collisions using the widely-adopted AMPT (A Multi-Phase Transport) model. The global angular momentum, the average vorticity carried by the quark-gluon plasma, and the locally defined vorticity fields are computed for Au+Au collisions, with detailed information of their time evolution, spatial distribution, as well as the dependence on beam energy and collision centrality.
\end{abstract}
\maketitle

\section{Introduction}

In relativistic heavy ion collisions a hot deconfined form of QCD matter, the quark-gluon plasma (QGP), has been created~\cite{Gyulassy:2004zy,Shuryak:2004cy}. In such collision experiments  at the Relativistic Heavy Ion Collider (RHIC) and the Large Hadron Collider (LHC), the QGP is found to undergo strong collective expansion as a relativistic fluid with extremely small dissipation~\cite{Heinz:2009xj}.

Recently there have been significant interests in the rotational aspects of the QGP collective motion, particularly regarding possible observable consequences of such rotation. Indeed in the non-central heavy ion collisions, there is a nonzero total angular momentum $J \propto b \snn$ (with $b$ as the impact parameter) carried by the system of two colliding nuclei.
Note that the beam energy $\snn$ is the nucleon-nucleon center-of-mass energy.
After the initial impact, most of this total angular momentum is carried away by the so-called ``spectators'' but there is a sizable fraction that remains in the created QGP and implies a nonzero rotational motion in the fluid. It was proposed a while ago that such rotation may affect the spin polarization of certain hadron production~\cite{Liang:2004ph,Becattini:2007sr}. More recent ideas  concern possible anomalous transport effects in a chiral QGP (for reviews and further references on this topic, see e.g. \cite{Kharzeev:2015znc,Liao:2014ava,Liao:2016diz}). The initial interest focused on effects induced by external electromagnetic
fields~\cite{Bloczynski:2012en} such as the well known Chiral Magnetic Effect, Chiral Magnetic Wave,
etc~\cite{Kharzeev:2004ey,Kharzeev:2007tn,Kharzeev:2007jp,son:2004tq,Metlitski:2005pr,Huang:2013iia,Burnier:2011bf}. It was later pointed out~\cite{Kharzeev:2007tn} that fluid rotation bears a lot of similarity to an external magnetic field and can also induce similar anomalous transport effects. One example is the Chiral Vortical Effect~\cite{Kharzeev:2007tn,Son:2009tf,Kharzeev:2010gr} which predicts a baryon current induced along the fluid rotation axis that can be measured via baryon separation across the reaction plane. Another example is the Chiral Vortical Wave~\cite{Jiang:2015cva} which predicts a baryonic charge quadrupole formed along the fluid rotation axis that can be measured via baryon/anti-baryon elliptic flow splitting.  Active experimental efforts are underway to detect possible signals of these effects, and it is of great phenomenological importance to quantify the rotational motion of the QGP in these collisions.

In this paper, we will present the quantification of QGP rotation in the relativistic heavy ion collisions, utilizing the tool of AMPT (A Multi-Phase Transport) model simulations. We will report our results for  the QGP global angular momentum, the average vorticity carried by the QGP, and the locally defined vorticity fields with detailed information of their time evolution,  spatial distribution,  as well as the dependence on beam energy and collision centrality. The rest of the paper is organized as follows: we give some general discussions on the fluid rotation in Section II; a brief discussion is given in Section III on our method of extracting rotational motion from AMPT and we further present results for the QGP angular momentum; we report results for  the vorticity fields and the fireball-averaged vorticity in Section IV; finally the summary is given in  Section V.

\section{Discussions on the Fluid Rotation}

\subsection{Angular Momentum and Vorticity}

The global rotation of a fluid can be quantified by the total angular momentum. For a many-body system of discrete classical (quasi)particles, one could calculate the total angular momentum $\vec J$ unambiguously by summing each particle's contribution together.
\begin{eqnarray}
\vec{J}=\sum_i \vec{r}_i \times\vec{p}_i
\end{eqnarray}
with $\vec{r}_i$ and $\vec{p}_i$ the position and momentum of each particle in given reference frame. For a large enough system after proper coarse-graining (e.g. like the fluid being made of many fluid cells), it can be considered as a continuous medium characterized by a series of locally defined quantities like momentum density, energy density and particle number  density $\vec{p}(\vec{r})$, $\epsilon(\vec{r})$ and $n(\vec{r})$ respectively. One then could rewrite the total angular momentum as
\begin{eqnarray}
\vec{J}&&=\int d^3r  \, \vec{r}\times\vec{p}(\vec r).
\end{eqnarray}

The fluid vorticity $\vec \omega$ is a more subtle quantity that is locally derived from local velocity field $\vec v(\vec r)$.  In the above coarse-graining picture, one may define the velocity field through the momentum and energy densities as  $\vec{v}(\vec r)=\vec{p}(\vec r)/\epsilon(\vec r)$ at each point/cell.  To avoid ambiguity, we will adopt  the familiar non-relativistic definition as
\begin{eqnarray} \label{eq_nr_vorticity}
\vec\omega=\nabla\times\vec{v}
\end{eqnarray}
In the case of a rigid body rotation with a global angular velocity around an axis, the above non-relativistic definition implies that the vorticity is  twice the rotational angular velocity. Of course a rotating fluid is quite different from a rigid body and in general the vorticity field is not constant across the fluid. It may be noted that in the relativistic hydrodynamic framework, a number of different quantities related to vorticity are often discussed in the literature~\cite{Becattini:2015ska}, such as the relativistic vorticity $\omega_\mu=-\frac{1}{2}\epsilon_{\mu\rho\sigma\tau}\omega^{\rho\sigma}u^{\tau}$, the T-vorticity, and the thermal vorticity. To the leading order non-relativistic expansion in fluid velocity, they all carry qualitatively the same information as the above defined $\vec \omega$.

The relation between angular momentum and the vorticity in general is rather complicated, and many factors could contribute to the angular momentum. For example the inhomogeneous distribution of energy density (i.e. inertia) could be a cause of nonzero angular momentum. Consider for example a situation with the whole system moving at the same velocity but with more matter located on one side than the other: the angular momentum will be nonzero indeed even without vorticity. But what we are interested in is the angular momentum associated with a nonzero vorticity. Let us use a simple example to examine the relation between the two. Consider a fluid (in some volume $V$) with a nonzero constant vorticity $\vec \omega$ (along certain rotation axis) and the corresponding flow field $\vec v= \frac{1}{2}\vec{\omega} \times \vec r$.  Noting that $\vec p = \epsilon \vec v$, the angular momentum  carried by the bulk of this fluid due to vorticity is given by:
$
\vec{J} =\int_V d^3r \, \vec{r}\times\vec{p}= \int_V d^3r \, \epsilon(\vec{r}) \,  \vec{r} \times \vec{v}  = \frac{1}{2}\int_V d^3r \, \epsilon(\vec{r}) \,  \vec{r} \times (\vec \omega \times \vec r) = \frac{1}{2} \int_V d^3r \, \epsilon(\vec{r})  [\vec r^2 \vec \omega - (\vec \omega \cdot \vec r) \vec r ]
$.
If the system is symmetric around the rotational axis $\hat{\omega}$, then the expression can be further simplified into
$
\vec{J} = \frac{1}{2} \int_V d^3r \, \epsilon(\vec{r})  [\vec r^2  - (\hat{\omega} \cdot \vec r)^2 ] \vec \omega = \frac{1}{2}\int_V d^3r \, [\rho^2 \epsilon(\vec{r}) ] \vec \omega
$
where $\rho^2= [\vec r^2  - (\hat{\omega} \cdot \vec r)^2 ]=r^2 [1 - (\hat{\omega}\cdot \hat{r})^2]$ is the distance-squared of point $\vec r$ from the rotational axis. Clearly the combination $[\rho^2 \epsilon(\vec{r})]$ plays the role of a sort of measure for the local fluid ``moment of inertia'' density.

 In a non-central heavy ion collision it is easy to see that the total angular momentum and the average vorticity (over event average) is  along the out-of-plane direction. Following usual convention we denote the beam direction as $\hat{z}$-axis, the impact parameter direction as $\hat{x}$-axis, while the out-of-plane direction as $\hat{y}$-axis.
A very useful quantity, that may be more directly related to the global rotation, is the fluid-averaged vorticity component along the $\hat{y}$-axis, which can be defined as
\begin{eqnarray}
\label{mean}
\langle \omega_y \rangle= \frac{ \int d^3 \vec r\ [{\cal W}(\vec r)] \,  {\omega}_y(\vec{r})}{ \int d^3 \vec r\ [{\cal W}(\vec r)]}
\end{eqnarray}
Note that in the above average we need proper local weighing function ${\cal W}(\vec r)$, for which we have chosen to use ${\cal W}(\vec r)={\rho}^2 \epsilon(\vec{r})$ (with $\rho$ the distance of point $\vec r$ to the $\hat{y}$-axis): this choice is motivated by the role of $[{\rho}^2 \epsilon(\vec{r})]$ as a sort of ``moment of inertia'' density.  Of course a different choice of the weighing function would lead to a different value for the average vorticity,
and we have investigated such uncertainty in this study.

\subsection{Results from A Simple Hard Sphere Model}

To get a qualitative and intuitive idea of the rotational aspect of heavy ion collisions, let us use the simplest model, the hard sphere model,  to estimate the total angular momentum. In this model, one  treats two heavy ions as two uniform 3D hard spheres (highly Lorentz-contracted along beam axis)
and calculate the angular momentum which remains in the overlapping zone (that supposedly represents  the fireball created in such collisions). In this model, the nucleus in its rest frame is a sphere of radius $R$ with homogenous number density. The radius can be estimated by  $R=R_0 \, A^{1/3}$ with $R_0=1.1 \rm fm$.
For a point $(x,y)$ on transverse plane in the overlapping zone, let us denote its distance to the two nucleus centers as $r_\pm \equiv \sqrt{y^2+(x\pm b/2)^2}$ respectively where $b$ is the impact parameter of the collision.
Thus the angular momentum along the y-axis is   given by
\begin{eqnarray}
&&J_y=\frac{A\,  \snn}{4\pi R^3/3}   \int x\ dx dy\,   \Theta\left(R-r_-\right) \, \Theta\left(R-r_+\right)  \nonumber\\
&&\times \left[ \sqrt{R^2-r_-^2}-\sqrt{R^2-r_+^2} \right] \,\, .
\end{eqnarray}
 Note that the integration is
performed only over the overlapping zone of the two nuclei. The sign of $J_y$ depends on the specific setup of coordinate axes and carries no specific meaning: for simplicity we take the convention such that the $J_y$ is positive.

In this simple hard sphere model, the angular momentum $J_y$ grows linearly with beam energy. Its dependence on the impact parameter is nonlinear: with increasing $b$, the distance between momentum-carrying nucleons and the center would increase (implying more contribution to angular momentum) while on the other hand the overlapping zone shrinks. One would expect a non-monotonic behavior for the $J_y$ dependence on $b$, with certain optimal impact parameter where the $J_y$ would be the largest.  Obviously this model is quite an oversimplified one. For example, the uniform density distribution is a rather crude approximation. Also the scatterings (especially that for  spectators) are totally ignored. Therefore, one may expect the hard sphere model to provide a qualitative, albeit not quantitatively accurate picture for the rotation. We will use it as a useful ``baseline'' for comparison with results from more quantitative and realistic simulations.

 Before presenting detailed results, let us mention in passing a number of studies on fluid rotation of heavy ion collisions in the literature. In \cite{Becattini:2007sr} angular momentum has been studied both with the hard sphere model and an intuitive hydrodynamical model.
It has been found that a more realistic distribution, the Woods-Saxon distribution, for nuclei would enhance the
fireball's angular momentum. The vorticity and angular momentum's effects on the emitted hadrons' polarization
have also been discussed. Some more detailed hydrodynamical simulations on the vorticity have been done in \cite{Becattini:2013vja,Csernai:2013bqa,Csernai:2014ywa,Becattini:2015ska}. Using Particle in Cell method,
it was found that for Pb-Pb collisions at $\snn=2.76$ TeV and $b=11$ fm the averaged vorticity
will start from around $0.14$ fm$^{-1}$ and drop to $0.015$ fm$^{-1}$ at around $7$ fm/c time.
Using the ECHO-QGP numerical code in \cite{Becattini:2015ska} the $\snn=200$ GeV Au-Au collisions
have been studied and found to have  a vorticity of the order of some $10^{-2}$ fm$^{-1}$   at freeze-out.
In the rest of this paper, we will report quantitative and detailed studies   for angular momentum as well as vorticity by using a quite different simulation tool, namely A Multi-Phase Transport (AMPT) model.

\section{Angular Momentum from AMPT Model Calculations}

\subsection{Setup of the AMPT Model}

Simulations of heavy ion events in this study were performed with A Multi-Phase Transport (AMPT) model \cite{Lin:2004en}.
The string melting version of the AMPT model \cite{Lin:2001zk,Lin:2004en} includes the initial particle production right after the primary collision of the two incoming nuclei, an elastic parton cascade, a quark coalescence model for hadronization, and a hadronic cascade.
Since our focus is the rotation of the quark-gluon plasma, we study the partonic matter with
the string melting version of AMPT, while effects of hadronization and hadron cascade on the rotation are not considered.
We use the same parameters as a previous study \cite{Lin:2014tya}, where it was shown that those parameters
reasonably reproduced the yields, transverse momentum spectra and $v_2$ data for low-$p_T$ pions and kaons in central and mid-central Au+Au collisions at $\snn=200$ GeV.
In particular, the parameters include the Lund string fragmentation parameters ($a=0.55$, $b=0.15$/GeV$^2$), strong coupling constant $\alpha_s=0.33$ for the parton cascade, a parton cross section of 3 mb (i.e. a parton Debye screening mass $\mu=2.265$/fm), and an upper limit of 0.40 on the relative production of strange to nonstrange quarks.

The AMPT model defines the impact parameter axis as the $x$-axis and the beam axis as the $z$-axis, where the incoming nucleus centered at $x=b/2>0$ ($x=-b/2<0$) has positive (negative) longitudinal momentum; this then creates an initial total angular momentum mostly along the (out-of-plane) $y$-direction.
A fraction of this angular momentum resides in the partonic matter created in the overlap region, while the rest of the angular momentum is essentially all due to spectator nucleons and is not included in the results shown in this study.
Note that all partons, regardless of their formation times, are included in calculating the fireball angular momentum or velocity field at a given time; this way the total angular momentum of the fireball is conserved throughout the fireball evolution.

\begin{figure}
\includegraphics[scale=0.8]{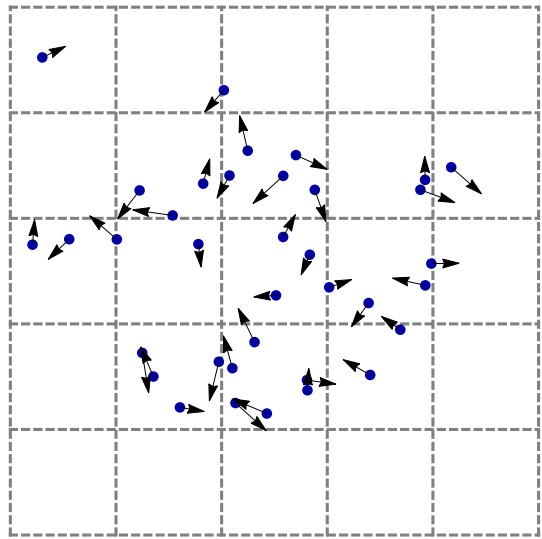}
\caption{Schematic picture for an AMPT event.}
\label{mesh}
\end{figure}

An advantage of the  AMPT model is that one  could explicitly track every parton or hadron's position and momentum at any given time. Fig. \ref{mesh} shows a schematic particle distribution
at a certain time where an arrow represents the particle's momentum direction.
In order to obtain a continuous momentum profile we first chose a proper volume, which is large enough to contain all of the particles and not too
large to have too many vacant areas. Then we divided it into small cells with proper size.
Using too large cells will conceal details, while using too small ones will enlarge the fluctuation in
the finite differential process.  The position of each cell, which could be treat as a fluid element, is indexed by the center, while its momentum and energy is defined by summing that of all the particles in it. This way we obtained the energy and momentum distribution for each event at a fixed time. Each event in the AMPT simulation would generate thousands of particles. This amount is still not large enough to generate a smooth momentum and energy distribution. It is necessary to generate thousands of events with the same collision energy and impact parameter. Finally we would get much smoother distributions by averaging these thousands of events.

Once we get distributions of momentum and energy, the velocity field could be defined as their ratio $\vec{p}/\epsilon$ for each cell. The angular momentum can be computed from directly summing over contributions from each individual cell. One could  calculate the vorticity in Eq.(\ref{eq_nr_vorticity}) with finite differential
method. For those vacant cells we would set their velocity as zero. This would generate large fluctuations for vorticity at the edge of the system (the interface between cells with a few particles and vacant cells). Such fluctuations are suppressed when computing the average vorticity because we adopt the energy density as the averaging weight as in Eq.(\ref{mean}). In our calculation we
chose the whole volume as $20\text{fm}\times 20\text{fm}$ on the transverse plane over a spatial rapidity span of $8$ units. Each cell's size is
$0.8\text{fm}\times 0.8\text{fm}$ on the transverse plane over a rapidity slice of $ 0.4$ unit. We have chosen the time step to be 0.2 fm/c for the vorticity analysis, and we analyze the parton matter up to the time of 9 fm/c in the center-of-mass frame.

\subsection{Angular Momentum of the QGP: Its Dependence on Time, Energy and Centrality}

We now present the results from AMPT for the angular momentum carried by the QGP fireball with detailed information on its time evolution as well as beam energy and collision centrality dependence.  Again the sign of $J_y$ depends on the specific setup of coordinate axes and carries no specific meaning: while the raw results from AMPT (due to its particular choice in the code) have negative sign, for simplicity we will just show results for the magnitude of $J_y$.

\begin{figure}
\includegraphics[scale=0.5]{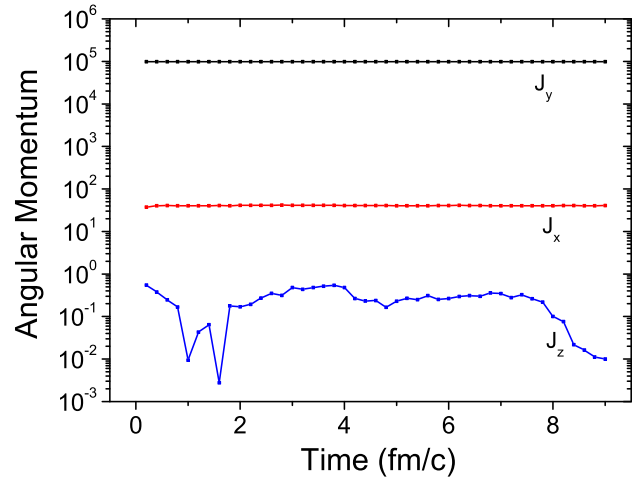}
\caption{Angular momentum from the AMPT model at $b=7$ fm and $\snn=200$ GeV.}
\label{amt}
\end{figure}

Let us first examine the time dependence of all the angular momentum components $J_{x,y,z}$ for given collision energy and centrality: see Fig. \ref{amt}. The results confirm the intuitive picture that the dominant component is $J_y$ (which is larger by orders of magnitude than $J_{x,z}$), i.e. the QGP global rotation is indeed around the out-of-plane axis. We also note that  the $J_y$ carried by the QGP fireball is about $10\sim20\%$ of the total angular momentum of the whole colliding system $J=A b\snn/2$. Lastly, $J_y$ is essentially a constant in time as it should be, which serves as a check of  the simulation's precision. These features are found to be the case for all other centralities as well as beam energies in our calculations.

\begin{figure}
\includegraphics[scale=0.5]{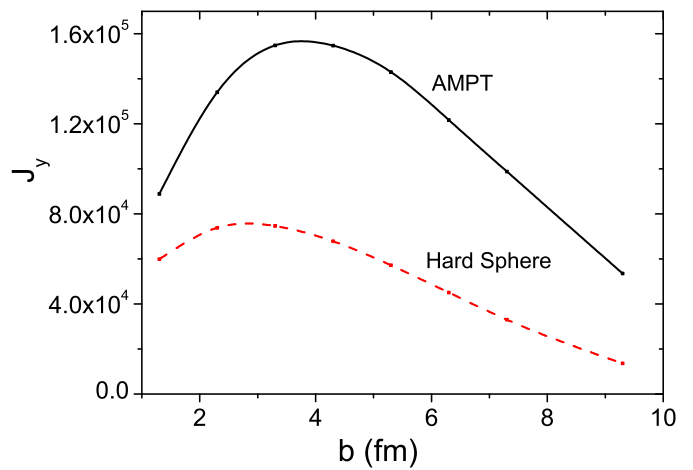}
\caption{Angular momentum $J_y$ as a function of $b$ from the AMPT model and the hard sphere model at $\snn=200$ GeV.}
\label{amb}
\end{figure}

\begin{figure}
\includegraphics[scale=0.5]{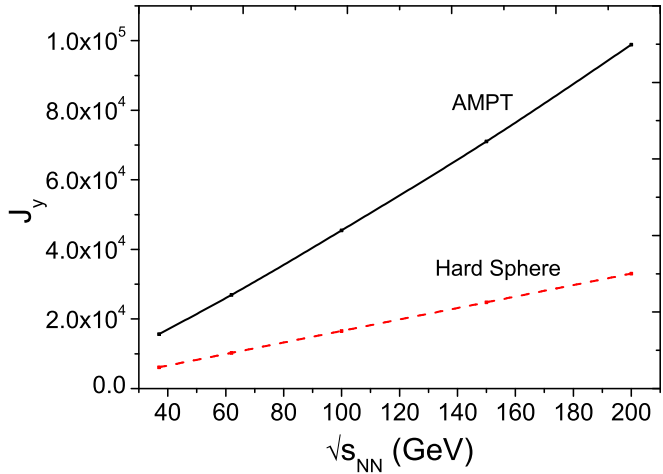}
\caption{Angular momentum $J_y$ as a function of $\snn$ from the AMPT model and the hard sphere model at $b=7$ fm.}
\label{ams}
\end{figure}

We next take a look at the dependence of $J_y$  on the collision energy and impact parameter in comparison with the results from the simple hard sphere model. Fig. \ref{amb} shows a non-monotonic dependence of $J_y$ on $b$ as expected, with a maximum around $b\sim 4$ fm. Fig. \ref{ams} shows an approximately linear growth of $J_y$ with increasing $\snn$, again as expected. In both figures, the $J_y$ from AMPT is about $2\sim 3$ times that from the hard sphere model.
Also note that the $b$ value corresponding to the peak in $J_y$ is also bigger from the AMPT model.
This can be understood from two factors. Firstly compared with the hard sphere model with sharp edges, the actual incident nuclear profile (Woods-Saxon in AMPT) is more extensive thus making the overlapping zone (where fireball is created) bigger, with more momentum carriers further away from the rotational axis at the center.  Secondly, in the hard sphere model the momentum carried outside the overlapping zone is not counted, while in actual collision (as captured by AMPT) the nucleons outside the geometric overlapping zone would still have probability to experience collision and become part of the fireball thus contributing more to the angular momentum.

\section{Vorticity from the AMPT Model }

\subsection{Local Vorticity Distribution}

Once the velocity distribution is obtained as described above, we can then use the finite differential method to calculate the vorticity numerically. We will focus on the vorticity along the out-of-plane direction, $\omega_y$. As this is a local quantity determined on each point/cell of the 3-dimensional space, we will first examine its distribution patterns over these spatial coordinates, with $x,y$ being transverse coordinates and $\eta$  the longitudinal spatial rapidity. All results in this subsection are for the case of Au+Au collisions at $b=7$ fm and $\snn=200$ GeV at   $t=1$ fm/c time. We have checked that computations performed with other parameters show the same patterns.


\begin{figure}
\includegraphics[scale=0.8]{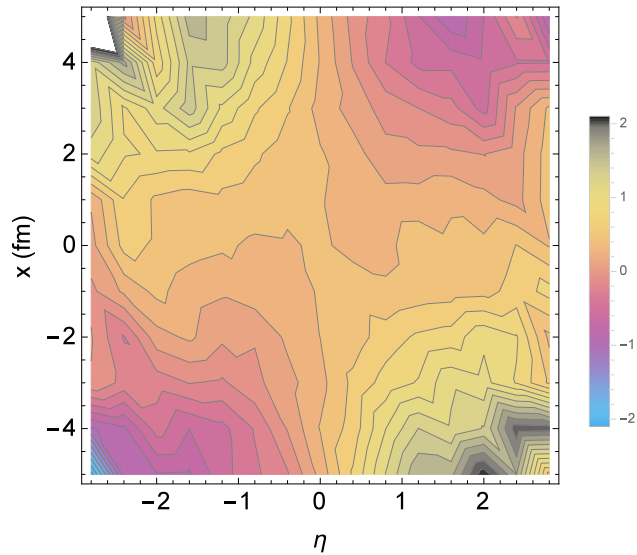}
\caption{$\omega_y$ (in the unit of fm$^{-1}$) profile at $y=0$ and $t=1$ fm/c with $b=7$ fm and $\snn=200$ GeV.}
\label{oxz}
\end{figure}

Fig. \ref{oxz} shows $\omega_y$ on the reaction plane (i.e. $y=0$). A qualitative pattern is observed to be as follows:  around the center of the fireball ($x\simeq 0$ or $\eta\simeq 0$), the $\omega_y$ is nearly vanishing; on the positive and negative side of $x$ or $\eta$ axes, the $\omega_y$ shows opposite sign, i.e. the vorticity is roughly an odd function of $x$ and $\eta$. A similar pattern has also been seen in hydrodynamic simulations e.g. in  \cite{Becattini:2015ska}.

\begin{figure}
\includegraphics[scale=0.8]{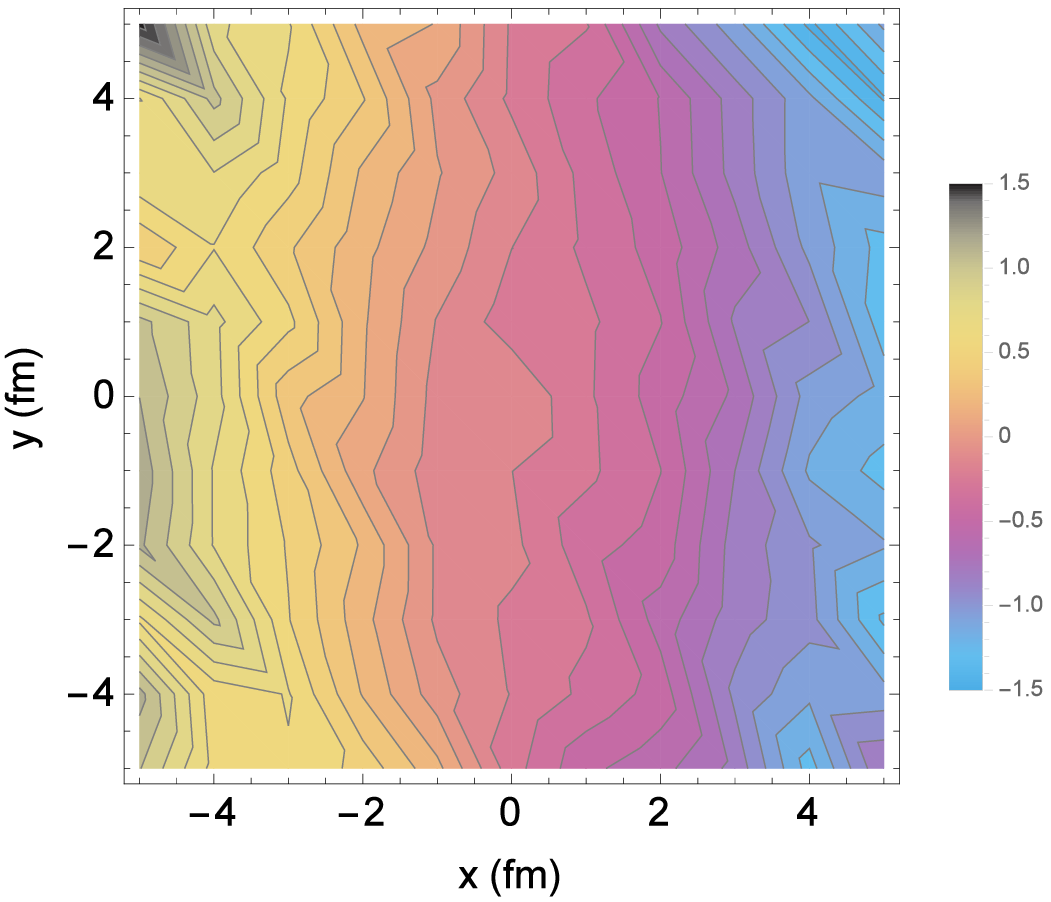}
\caption{$\omega_y$ profile at $\eta=1$ and $t=1$ fm/c with $b=7$ fm and $\snn=200$ GeV.}
\label{avevoxy1}
\end{figure}

\begin{figure}
\includegraphics[scale=0.8]{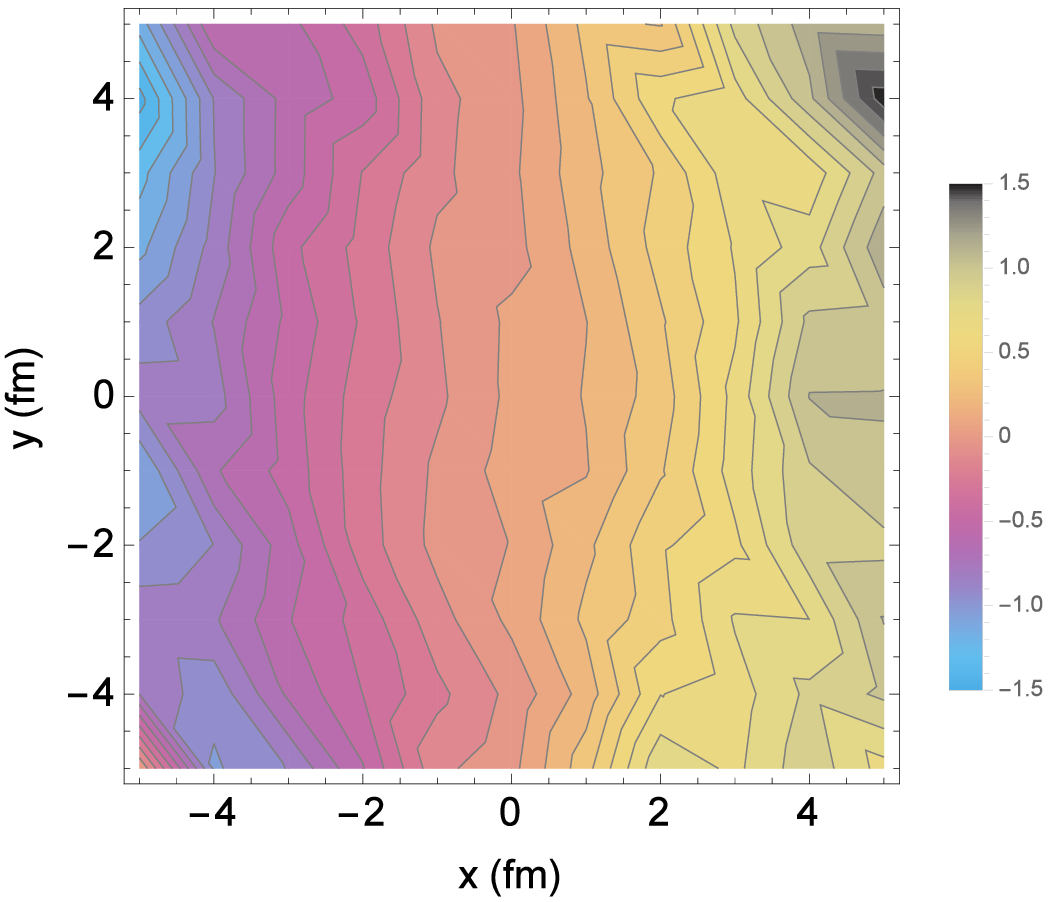}
\caption{$\omega_y$ profile at $\eta=-1$ and $t=1$ fm/c with $b=7$ fm and $\snn=200$ GeV.}
\label{avevoxy2}
\end{figure}

Figs. \ref{avevoxy1} and \ref{avevoxy2}  show the profile of vorticity  on the transverse plane, at forward and backward rapidity $\eta = \pm 1$ respectively. These two profiles again demonstrate the $\omega_y$ as roughly an odd function of $x$ and $\eta$. The dependence of $\omega_y$ on $y$ appears rather mild and roughly symmetric between positive and negative $y$ directions. An important observation is that the region at large $x$ values has the biggest $\omega_y$.

\begin{figure}
\includegraphics[scale=0.8]{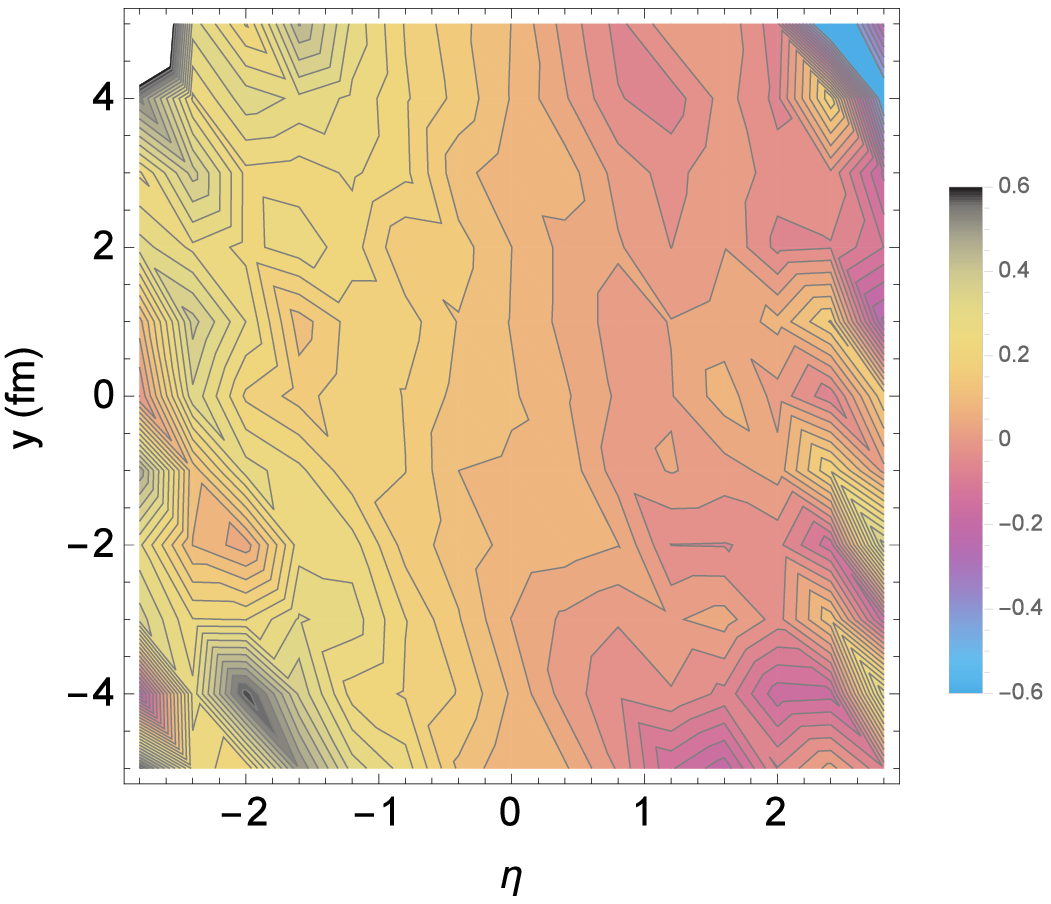}
\caption{$\omega_y$ profile at $x=1$ fm and $t=1$ fm/c  with $b=7$ fm and $\snn=200$ GeV.}
\label{avevoyz2}
\end{figure}

\begin{figure}
\includegraphics[scale=0.8]{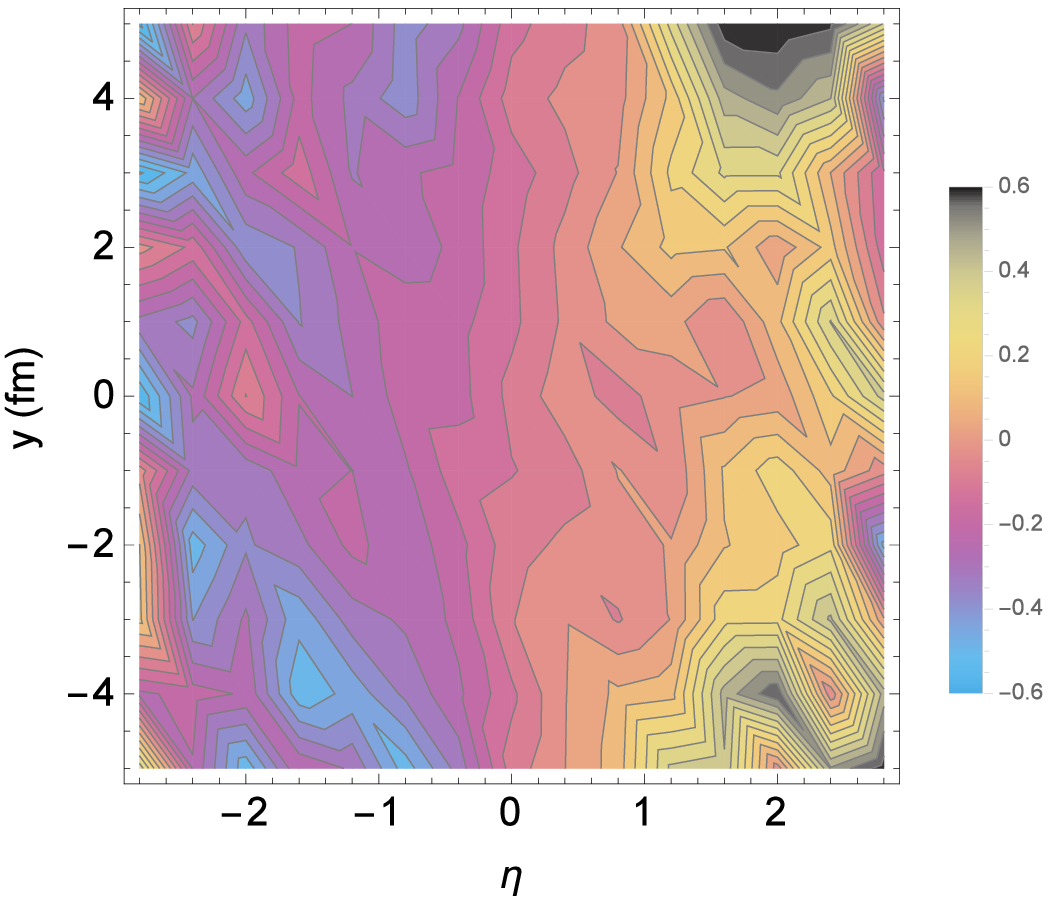}
\caption{$\omega_y$ profile at $x=-1$ fm and $t=1$ fm/c  with $b=7$ fm and $\snn=200$ GeV.}
\label{avevoyz1}
\end{figure}

Finally in Figs. \ref{avevoyz1} and \ref{avevoyz2}  we show the vorticity profile on the $y-\eta$ plane for $x=\pm 1$ fm. Again one sees a pattern that is consistent with what we've seen from the plots of $x-y$ and $x-\eta$ profiles.


\begin{figure}
\includegraphics[scale=0.8]{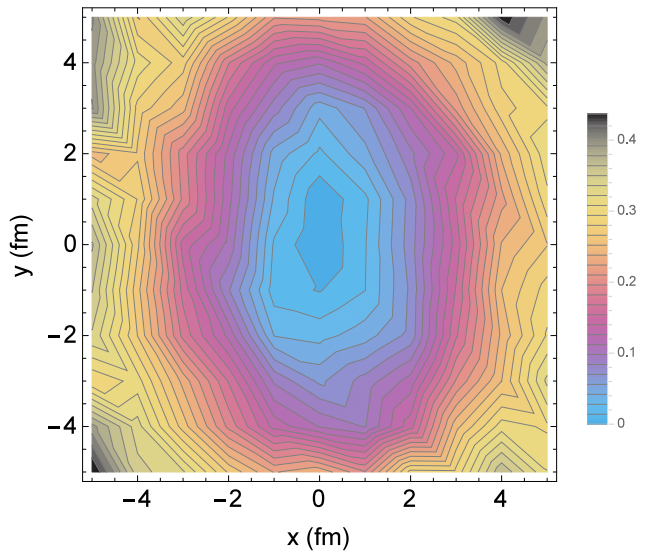}
\caption{Radial velocity profile at $\eta=1$ and $t=1$ fm/c with $b=7$ fm and $\snn=200$ GeV.}
\label{vrxy}
\end{figure}

Given these observed patterns, it is tempting  to ask: what is the origin of these particular patterns, and how are they related (or unrelated) to the global rotation that we are interested in.  As we already pointed out in the previous section, a locally nonzero vorticity is not directly linked with a global angular momentum. Indeed  as it turns out, most of these patterns could be understood simply from the underlying radial flow that is unrelated to the rotational motion. To see this, let us consider a radial flow profile $\vec v$ that can be well parameterized in the following form:
\begin{eqnarray}
\vec{v}(\rho,\phi,\eta)&=& \hat{e}_\rho v_0(\rho, \eta) \left [ 1+  2c_2(\rho, \eta)\cos 2\phi \right ] \qquad
\end{eqnarray}
where $\hat{e}_\rho$ is the unit vector along the transverse radial direction, $\rho$ and $\phi$ are transverse radial and azimuthal coordinates.
 Fig.\ref{vrxy}   shows the transverse radial velocity profile that we have extracted from the same AMPT simulations. We have checked that the velocity along $\vec{e}_\phi$ is negligibly small as compared with   the radial component. One can compute the local vorticity $\omega_y$ resulting solely from the above velocity profile using the definition in Eq.(\ref{eq_nr_vorticity}):
\begin{eqnarray}
\omega_y&&=\frac{\partial v_\rho}{\partial z}\cos\phi\nonumber\\
&&=\frac{1}{t}(ch\eta)^2\ \partial_\eta (v_0+ 2v_0c_2\cos 2\phi)\cos\phi\nonumber\\
&&=\frac{1}{t}(ch\eta)^2\ \left(\frac{x}{\rho}\right)\ \partial_\eta \left [v_0+2v_0c_2(2\frac{x^2}{\rho^2}-1) \right]
\end{eqnarray}
As many hydrodynamic simulations assume, and as indeed confirmed in our AMPT simulations, the coefficients $v_0$ and $c_2$ are both even functions of $\eta$ to a very good approximation.
As such both terms, $\partial_\eta v_0$ and $\partial_\eta(v_0 c_2)$, are odd functions of $\eta$. Therefore the sole contribution to $\omega_y$ from radial flow is indeed approximately an odd function of both $x$ and $\eta$. The dependence of $\omega_y$ on $y$ is only through $\rho=\sqrt{x^2+y^2}$ which is indeed even and mild function of $y$. Therefore, we see that most of the qualitative patterns of the $\omega_y$ profiles from AMPT computations can be reasonably understood from the radial flow contribution. Furthermore, we have checked quantitatively that indeed the values of $\omega_y$ are dominated by such contributions.

So what does that imply? It suggests that, to extract the component of local vorticity that is truly associated with the global rotation, one needs to perform an average over the fireball. Upon such averaging, the background flow contributions to local vorticity would cancel out, and what remains can be attributed to the rotational motion.

\subsection{Averaged Vorticity for the QGP}

In this subsection we present our key results: the properly averaged vorticity $\langle \omega_y \rangle$ that encodes information on the  global rotation of the fireball. For the averaging process,  we will use the weighing function as given in Eq.(\ref{mean}) for the fireball over the full transverse plane and a spatial rapidity span of $\eta\in [-4,4]$.

\begin{figure}
\includegraphics[scale=0.45]{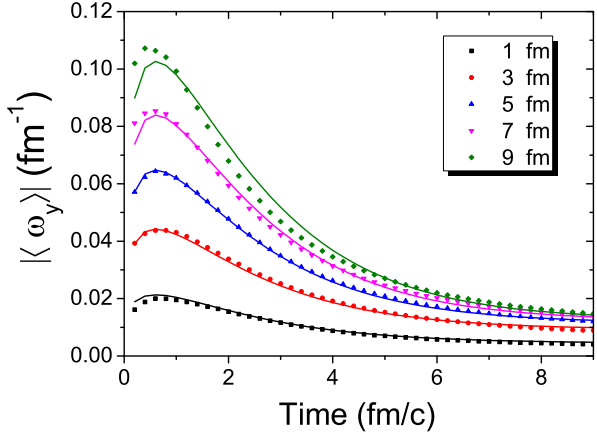}
\caption{Averaged vorticity $\langle \omega_y \rangle$  from the AMPT model as a function of time at various impact parameter $b$ for fixed beam energy $\snn=200$ GeV. The solid curves are from fitting formula (see text for details). }
\label{aveob}
\end{figure}

Let us first present the centrality dependence of $\langle \omega_y \rangle$ at given beam energy $\snn=200$ GeV: see Fig.~\ref{aveob}. The  $\langle \omega_y \rangle$ briefly increases with time which is most likely due to parton scatterings during the early stage (when the transverse radial expansion is not developed yet) that in certain way decrease the fluid moment of inertia. The averaged vorticity reaches peak value at an almost universal time around $1$ fm/c and then follows a steady decrease with time. The decrease is due to the system's expansion which increases total moment of inertia at the price of reduced vorticity due to the constraint of constant angular momentum. The results also clearly demonstrate that the averaged vorticity increases from central to peripheral collisions: this trend is different from the angular momentum. Such difference again can be understood as follows: while the vorticity increases with $b$, the fluid moment of inertia (pertinent to rotation) in the fireball decreases with $b$, thus the angular momentum shows a non-monotonic behavior due to the two competing trends.

\begin{figure}
\includegraphics[scale=0.45]{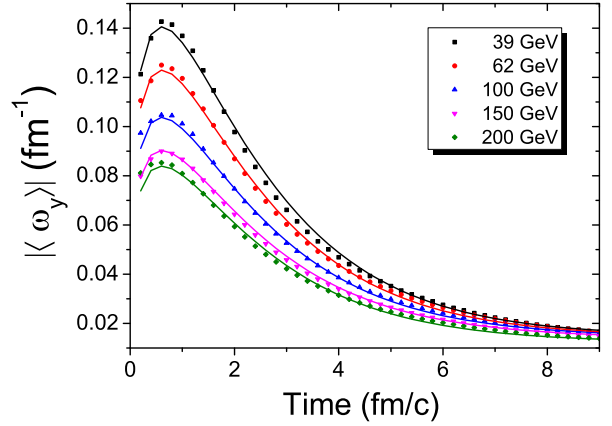}
\caption{Averaged vorticity $\langle \omega_y \rangle$  from the AMPT model as a function of time at varied  beam energy $\snn$ for fixed impact parameter $b=7$ fm. The solid curves are from fitting formula (see text for details). }
\label{aveos}
\end{figure}

We next show the beam energy dependence of $\langle \omega_y \rangle$ at given impact parameter $b=7$ fm: see Fig.~\ref{aveos}. Similar time evolution patterns are observed at all energies. We notice that the averaged vorticity increases with decreasing beam energy, in quite the opposite trend to the angular momentum. This may be understood as follows: with increasing beam energy, the fluid moment of inertia (pertinent to rotation) increases more rapidly than the decrease of vorticity, thus the total angular momentum is still increasing. We have numerically checked that this is indeed the case.

Finally, we present a parameterization of averaged vorticity as a function of time, centrality as well as beam energy, which provides comprehensive and very good fit to the numerical results of Au+Au collisions
from AMPT. This is given by:
 \begin{eqnarray}  \label{eq_omega_fit1}
 \langle \omega_y \rangle (t, b, \snn) &=&  A(b,\snn) \nonumber \\
  && + B(b,\snn) \left( 0.58 t \right)^{0.35} \, e^{-0.58 t} \qquad
 \end{eqnarray}
with the two coefficients $A$ and $B$ given by:
\begin{eqnarray}
A &=& \left[ e^{-0.016 \,b \,\sqrt{s_{NN}}} + 1 \right] \times  \tanh(0.28\, b) \nonumber \\
&& \,\, \times  \left[ 0.001775 \tanh(3-0.015\sqrt{s_{NN}}) + 0.0128 \right]  \,\, , \nonumber  \\
B &=&    \left[ e^{-0.016 \,b \,\sqrt{s_{NN}}} + 1 \right] \times  \left[ 0.02388 \, b + 0.01203 \right]  \nonumber \\
&& \,\, \times  \left[ 1.751 - \tanh(0.01 \sqrt{s_{NN}})  \right]  \nonumber
\,\, .
\end{eqnarray}
In the above relations, $\snn$ should be evaluated in the unit of GeV, $b$ in the unit of fm, $t$ in the unit of fm/c, and $\omega_y$ in the unit of fm$^{-1}$.
The solid curves in Figs.~\ref{aveob} and \ref{aveos} are obtained from the above formula, in comparison with actual AMPT results. As can be seen, the agreement is excellent and we have checked that in all cases the relative error of the above formula is at most a few percent. Such parameterization could be conveniently used for future studies of various vorticity driven effects in QGP.

\subsection{Study of Uncertainties}

In this last part, we investigate a number of uncertainties in quantifying the averaged vorticity.

\begin{figure}
\includegraphics[scale=0.45]{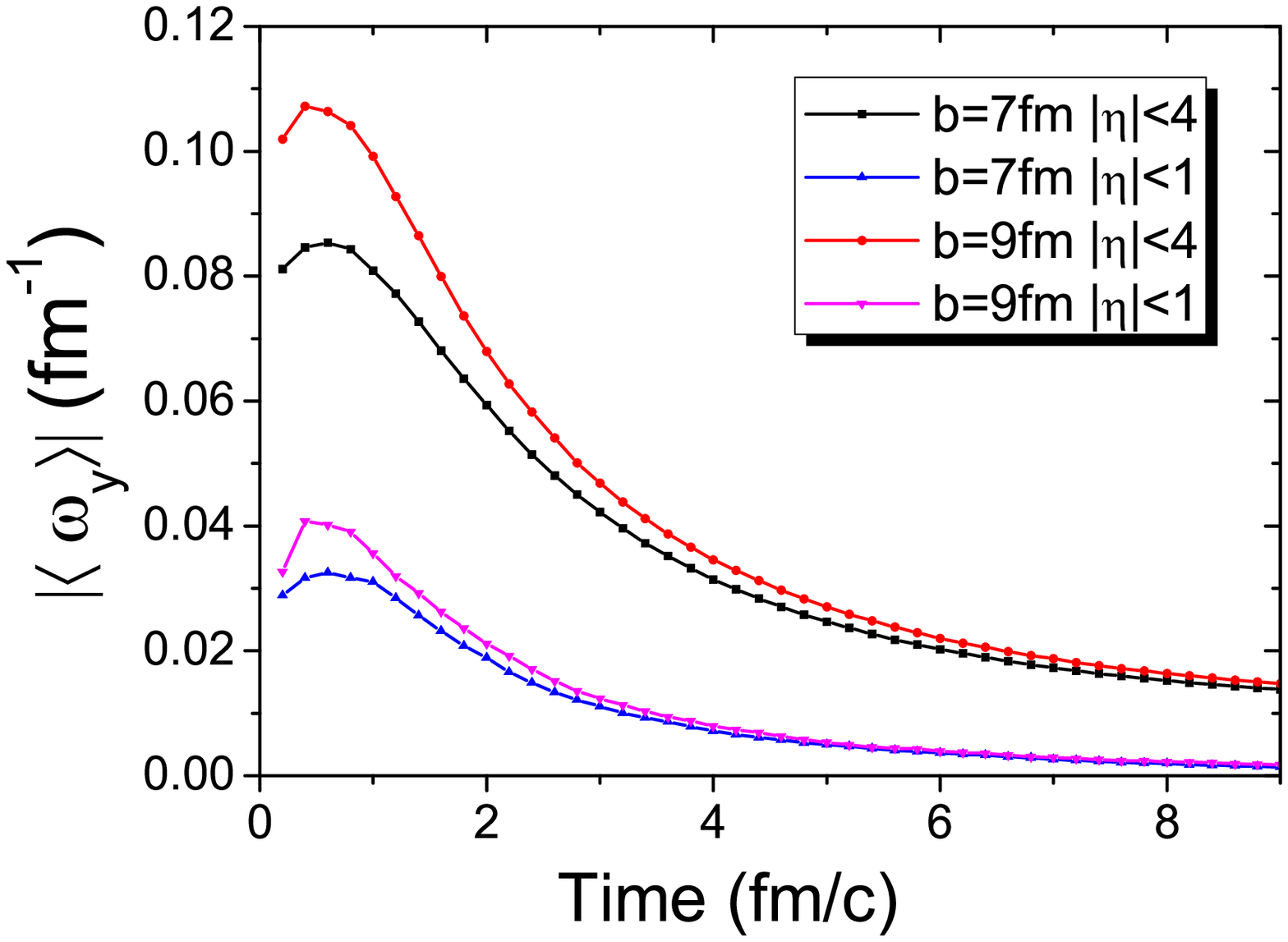}
\caption{Averaged vorticity $\langle \omega_y \rangle$, with spatial rapidity span $\eta\in [-1,1]$ and $\eta\in [-4,4]$ respectively,  from the AMPT model as a function of time at  $\snn=200$ GeV for fixed impact parameters $b=7,9$ fm.  }
\label{etapm1}
\end{figure}

\begin{figure}
\includegraphics[scale=0.45]{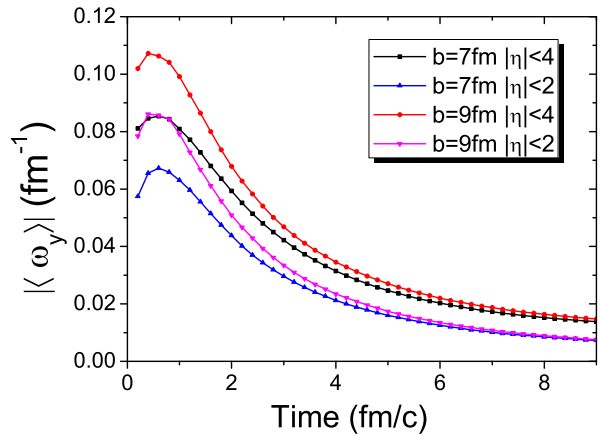}
\caption{Averaged vorticity $\langle \omega_y \rangle$, with spatial rapidity span $\eta\in [-2,2]$ and $\eta\in [-4,4]$ respectively,  from the AMPT model as a function of time at  $\snn=200$ GeV for fixed impact parameters $b=7,9$ fm.  }
\label{etapm2}
\end{figure}

 One uncertainty is related to the choice of volume in performing the average. In the previous subsection we have chosen to average over the spatial rapidity span of $\eta\in [-4,4]$. However when it comes to certain specific vorticity driven effects and the pertinent final hadron observables, it is not 100\% clear what is precisely the relevant longitudinal volume. To get an idea of this uncertainty, we have computed the $\langle \omega_y \rangle $ for different choices of spatial rapidity span: see Fig.~\ref{etapm1} for results from $\eta\in [-1,1]$ in comparison with those from $\eta\in [-4,4]$;  see Fig.~\ref{etapm2} for results from $\eta\in [-2,2]$ in comparison with those from $\eta\in [-4,4]$. As one can see from the comparison, at early to not-so-late time, the results differ by about a factor of two between $\eta\in [-1,1]$ and $\eta\in [-4,4]$ while differ by about 30\% percent or so between $\eta\in [-2,2]$ and $\eta\in [-4,4]$. At late time the results with $\eta\in [-4,4]$ are significantly larger than the others. Clearly the contributions to the averaged vorticity from large spatial rapidity region become dominant at late time. In the modeling of vorticity driven effects, such uncertainty needs to be carefully considered.  For the convenience of future applications, we also provide here the parameterization for averaged vorticity  computed with a spatial rapidity span of $\eta\in [-1,1]$:
 \begin{eqnarray}  \label{eq_omega_fit2}
 \langle \omega_y \rangle (t, b, \snn) &=&  A(b,\snn) \nonumber \\
&& + B(b,\snn) \left( 0.65 t \right)^{0.3} \, e^{-0.65 t} \qquad
 \end{eqnarray}
 where
 \begin{eqnarray}
A &=& \tanh(0.35\, b) \nonumber \\
&& \,\, \times  \left[ 0.00143 \tanh(1.5-0.015\sqrt{s_{NN}}) + 0.00271 \right]  \,\, , \nonumber  \\
B &=&    \left[ 0.0123 \, b + 0.0261 \right] \,\, \times  \left[ 1.42 - \tanh(0.008 \sqrt{s_{NN}})  \right]  \nonumber
\,\, .
\end{eqnarray}
Again in the above relations, $\snn$ should be evaluated in the unit of GeV, $b$ in the unit of fm, $t$ in the unit of fm/c, and $\omega_y$ in the unit of fm$^{-1}$.

Another uncertainty is related to the weighing function ${\cal W}(\vec r)$ used in the averaging process, see Eq.(\ref{mean}). While our choice of ${\cal W}(\vec r)={\rho}^2 \epsilon(\vec{r})$ is well motivated, one may still wonder to what extent the results for $\langle \omega_y \rangle$ may be specific to such choice. For comparison, we have computed the $\langle \omega_y \rangle$ with three other  choices for the weighing function: ${\cal W} \to {\rho}^2 n(\vec r)$,  ${\cal W} \to  n(\vec r)$, and ${\cal W} \to  \epsilon(\vec r)$ (where $n(\vec r)$ is simply the local parton number density in AMPT). One observes a factor of $2\sim 3$ variation among these different choices, and our original choice of ${\cal W} \to {\rho}^2 \epsilon(\vec r)$ gives the largest averaged vorticity. This comparison provides a reasonable idea of the degrees of uncertainty associated with the averaging process.

\begin{figure}
\includegraphics[scale=0.45]{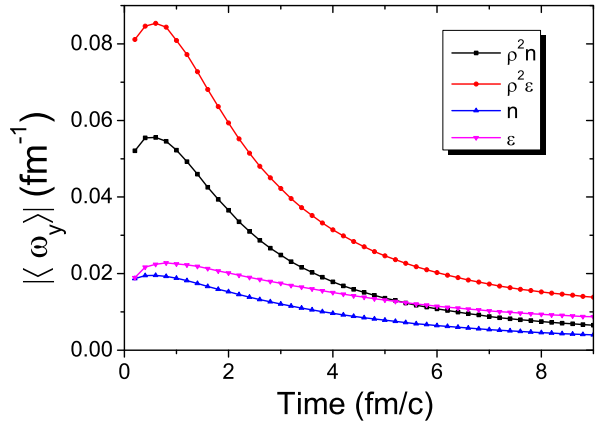}
\caption{Averaged vorticity $\langle \omega_y \rangle$   from the AMPT model as a function of time at  $\snn=200$ GeV for fixed impact parameter $b=7$ fm, with four different choices of weighing functions in performing the average (see text for details).   }
\label{weight}
\end{figure}

\section{Summary}

In summary, we have used the AMPT simulations to model relativistic heavy ion collisions and extract information on the rotational motion of the created quark-gluon plasma.
  In a general non-central collision, there is obviously a nonzero global angular momentum $J_y \propto b \snn$. While the majority of this angular momentum is carried away by the spectator nucleons, our computations have shown that a considerable fraction (about $10\sim 20\%$) of $J_y$ remains carried by the QGP in the collision zone and is essentially conserved in time. This implies a relatively long time duration of the global rotation that may drive interesting phenomena such as the Chiral Vortical Effect and the Chiral Vortical Wave. Apart from event-by-event fluctuations,   this angular momentum is on average   pointing  in the out-of-plane direction. We have also computed the  local vorticity field $\omega$ in the QGP fireball and analyzed its distributions over the transverse plane and spatial rapidity. We have identified the patterns of $\omega$ that come from the usual background collective flow (without rotation) and  have quantified properly averaged out-of-plane vorticity that is associated with the collective rotational motion. Detailed results for these important quantities' time evolution as well as their dependence on beam energy and collision centrality have been reported.
  Parameterizations for the numerical results of $\langle \omega_y \rangle$ as a function of time $t$, impact parameter $b$, and beam energy $\snn$ are provided, which could be a convenient tool  for future modelings of vorticity driven effects in heavy ion collisions.  To conclude, we expect this study to provide crucial input for efforts in the near future to quantify observable effects associated with the rotational motion of the quark-gluon plasma.

\section*{Acknowledgements}

The authors thank Huan Huang, Xu-Guang Huang, Dmitri Kharzeev, Aihong Tang, Fu-qiang Wang, Gang Wang, Nu Xu and Zhangbu Xu for helpful discussions. We also thank Hui Li and Qun Wang for pointing out the important typo related to Eq.(3) in a previous version of this paper. The research of YJ and JL is supported in part by the US National Science Foundation (Grant No. PHY-1352368) and by the US Department of Energy through the Beam Energy Scan Theory Collaboration. JL is also grateful to the RIKEN BNL Research Center for partial support, and acknowledges the Simons Center for Geometry and Physics, Stony Brook University at which some of the research for this paper was performed.

\end{document}